\documentclass[aps, prd, twocolumn, superscriptaddress, nofootinbib, longbibliography, 10pt]{revtex4-1}

\usepackage{amsthm}
\usepackage{esvect}
\usepackage{import}
\usepackage{graphicx}
\usepackage{dcolumn}
\usepackage{bm}

\usepackage[caption=false]{subfig}

\usepackage{braket}
\usepackage{slashed}
\usepackage{geometry}
\usepackage{multirow}
\usepackage{appendix}

\usepackage{cancel}
\usepackage{indentfirst}
\usepackage{hyperref}

\usepackage[utf8]{inputenc}

\graphicspath{{./}}

\usepackage{amsmath,setspace, amsfonts,latexsym}
\usepackage{amssymb}
\usepackage{color}
\usepackage{epsfig}

\usepackage{ulem}
\usepackage{array}
\usepackage[dvipsnames]{xcolor}

\hypersetup{pdfauthor={Subin Han}}

\DeclareMathOperator*{\jet}{\text{jet}}
\DeclareMathOperator*{\mcR}{\mathcal{R}}
\DeclareMathOperator*{\mcV}{\mathcal{V}}

\begin{document}

\title{Elaborating Higgs to dimuon decay from gluon fusion \\
by decorrelation and jet substructure}
\author{Subin Han}
\author{Hyung Do Kim}
\affiliation{Department of Physics and Astronomy and Center for Theoretical Physics, Seoul National University, Seoul, Korea 08826}

\begin{abstract}
Discovery of the Higgs boson decay to dimuon is anticipated soon based on the current evidence. Precise categorization of events without affecting the invariant mass shape is crucial in the analysis. Decorrelation of the invariant mass and the output of discriminators (the score of discriminators) is essential for consistent and precise analysis. In this paper, we use distance correlation as the additional loss function to achieve the decorrelation for discriminators and examine various analysis methods. The significance of the Higgs to dimuon signal from gluon fusion is considerably improved by including jet substructure variables. 
\end{abstract}

\maketitle
            \section{Introduction}
            After Higgs discovery \cite{Higgs_discovery_CMS, Higgs_discovery_Atlas}, various properties of Higgs have been measured including Higgs couplings to Standard Model (SM) particles. The SM predicts the Higgs coupling to a particle is proportional to its mass. Accordingly, Higgs couplings to heavy particles such as gauge bosons and third generation fermions are well measured \cite{PhysRevD.101.012002, Higgs_to_top_discovery, Higgs_to_bottom_discovery, Higgs_to_tau_discovery}. However, couplings to relatively light particles have not been satisfactorily measured so far. The Higgs to dimuon channel is the most sensitive channel among the second generation fermions at the LHC due to the relatively clean leptonic final states. The property of Higgs couplings to light fermions can verify the Higgs mechanism of the SM more accurately. Furthermore, the growing interest towards a muon collider makes this channel more interesting.
            
            In 2020, the decay of Higgs to dimuon was confirmed to be barely consistent with the SM \cite{CMS:HiggsToDimuon}. The observed (expected) significance was 3.0$\sigma$ (2.5$\sigma$). The analysis was performed separately for each Higgs production channel. As the dimuon invariant mass $m_{\mu\mu}$ is an excellent variable for distinguishing the resonance peak of the Higgs signal from the background events, the extraction of the number of signal and background events as well as the calculation of discovery significance were conducted by fitting data to the expected $m_{\mu\mu}$ distribution. Several uncertainties arising from the simulation could be reduced through shape analysis during the distribution fitting. However, the signal-to-background ratio for each channel was not sufficient to extract the number of signal events from $m_{\mu\mu}$ fitting($\sim 0.3\%$ for gluon fusion channel at Higgs mass). To increase the sensitivity, each channel was further divided into several categories with different signal-to-background ratios based on the output of discriminators. 
            
            A crucial requirement for the categorized invariant mass fitting is that the information about the invariant mass should not be used in the categorization process. In other words, the output of discriminators, or the discriminator score, should be statistically independent of the invariant mass. Categorization by a score which is statistically dependent on the invariant mass generally results in different invariant mass distributions for each category, $f(m | \text{score}) \neq f(m)$, where $f$ represents the probability density function. If the invariant mass distribution is distorted, the number of events extracted from each category can not be reliably trusted. In extreme cases, a resonance peak near the Higgs mass may appear in the background $m_{\mu\mu}$ distribution, adding significant uncertainties on the number of signal events near the Higgs bump region. In addition, if we fit to the invariant mass after the categorization by a discriminator correlated to the invariant mass, it results in double usage of the invariant mass, compounding the error. Therefore, the standard likelihood fitting to the invariant mass distribution can be consistently combined with machine discriminator results only when two results are decorrelated/deassociated. \footnote{Technically, the term `deassociate/deassociation' is more appropriate \cite{association_correlation_nature}, but we will use the term `decorrelate/decorrelation' following other literature}
            
            To avoid the distortions in the invariant mass distribution that arise when events are selected by a discriminator correlated with the invariant mass, researchers have not only securely excluded the invariant mass, but also carefully selected other variables to make the categories as insensitive to the invariant mass as possible. However, as will be seen in the paper, machine discriminators generally learn about invariant mass since it has a strong discrimination power. An ad hoc choice of variables which seemingly unrelated to the invariant mass, qualitative check of distortion in distributions after categorization \cite{ML_CMS} and a posteriori modification of fitting functions are not satisfactory solutions. In this paper we provide a quantitative and consistent decorrelation framework to deal with this problem.

            There are numerous techniques of decorrelation have been developed to decouple a discriminator from a specific feature \cite{Adversarial_pivot, Adversarial_jetsub, Adversarial_spannowsky, Adversarial_domain_adpatation, conditional_normalizing_flow, optimal_transport, DDT, CSS, mutual_info_MINE, MoDe, autoencoder, QBDT, ABCDisco, Atlas_decorr, CMS_decorr, Wunsch:2019qbo}. Among those various methods, we choose to use a measure of dependence `distance correlation' first used for decorrelation in High Energy Physics by \cite{Disco_fever}. The distance correlation is capable of capturing both linear and non-linear dependence, well-defined in both population and sample sapces, and differentiable, so being compatible with deep learning frameworks. At the same time, the range is well defined such that it is 0 when there is no correlation and is 1 when it is fully deterministic/dependent. By leveraging the flexiblility of deep neural network (DNN), we use the distance correlation between $m_{\mu\mu}$ and output of discriminator as an additional loss for training. As a result, it has been found that almost perfect statistical independence between $m_{\mu \mu}$ and the output of discriminator can be systematically achieved for both signal and background data classes simultaneously. It allows us to choose the required degrees of independence, which is crucial for accurate analysis, despite being in a trade-off relation to the performance of the discriminator. This approach ensures reliable categorization without compromising the integrity of the invariant mass distribution.

            Compared to the analysis using only the kinematic variables of reconstructed particle objects, incorporating jet substructure variables of the jets associated with the dimuon system as additional input variables can provide extra power, especially in discriminating the gluon dominant gluon fusion (ggH) signal from other background processes. In \cite{Cho_2020}, it has been demonstrated that the ggH channel can be exploited to improve the sensitivity of the Higgs invisible signal if jet substructure variables are fully utilized to distinguish the flavors and kinematics of initial state radiation (ISR) by DNN analysis. The differences in parton luminosities and QCD color factors of vertices make the ggH ISR to be gluon-rich, while make the background Drell-Yan (DY) ISR to be quark-rich. In addition, due to the helicity conservation, ggH ISR has a deficit of quark jets at the central rapidity region\footnote{There is no issue of decorrelation in Higgs to invisible study. The treatment of systematic uncertainties originating from the simulation remain as an open question.}.
            In the Higgs to dimuon discovery, currently the vector boson fusion (VBF) is the leading production channel despite the fact that the ggH production cross section is more than 10 times larger than VBF one at $\sqrt{s} = 13\ $TeV \cite{https://doi.org/10.23731/cyrm-2017-002}. Therefore we included the jet substructure variables to improve the sensitivity of Higgs to dimuon signal from ggH channel and compared the results.

            The paper is composed as following. 
            In section \ref{sec: analysis strategy}, we first outline the event simulation and ggH selection cuts. We also summarize the analysis strategy including training variables, common training schemes, and the categorization procedure.
            In section \ref{sec:discriminator}, we introduce three types of discriminators and compare their result in detail. In particular, with the $m_{\mu\mu}$ decorrelated discriminator implemented using the distance correlation, the degree of decorrelation and the discrimination performance will be compared.
            In section \ref{sec: discovery significance}, we compute the discovery significance using the categorized events and check the effect of including jet substructure variables.
            \section{Events and analysis strategy}\label{sec: analysis strategy}
        \subsection*{Event simulation}
        When an event is classified as being from ggH, it means that the event passed all selection cuts to identify ggH, regardless of its genuine origin. Therefore, all possible Higgs production channels should be simulated. However, the contribution of other channels to the identified ggH signals is much smaller than that of the genuine ggH channel, as their cross sections are much smaller. The VBF channel has the next-to-leading-order contribution which is less than 5\% of the ggH channel. Thus we simulated only the ggH and VBF channels. Similarly, there are many channels that contribute as backgrounds to ggH. Among them, we included DY, which is the dominant background and t$\Bar{\text{t}}$, which is more than 20\% of DY for multi-jet events. All the events were simulated by MadGraph5\_aMC@NLO \cite{Alwall_2014} at $\sqrt{s} = 13$ TeV and NLO in QCD. Each simulation was done inclusively for the number of jets, so that we can properly include both soft and hard jet emissions.

        \subsection*{Event selection}
        After the simulation, events are selected for ggH identification by the criteria as the following. Basically, an event is required to have two muons with opposite charges, each with $p_T > 20\ $GeV and $|\eta| < 2.4$. Any events including b-jet and/or additional charged lepton are vetoed to minimize overlap with t$\Bar{t}$H and VH channels. Finally, events with two energetic jets with a large separation in pseudorapidity ($m_{jj} > 400\ $GeV, $|\Delta \eta_{jj}| > 2.5)$ are rejected as these are more likely to be identified as VBF. 

        \subsection*{Training variables}
        To improve the signal-to-background ratio, various information other than $m_{\mu\mu}$ is used to divide the total data into some categories. Recently, with the development of machine learning techniques, output of machine learning discriminators is frequently used for categorization and allows us to use multiple variables simultaneously.

        The training variables used by the CMS collaboration are denoted by $S_{n_{\jet}}$ for each jet multiplicity $n_{\jet}$. $S_{n_{\text{jet}}}$ are carefully selected to exclude $m_{\mu\mu}$ information. See Appendix \ref{app: cms variables} for more details. Please note that $n_{\text{jet}} = 2$ means events with at least two jets, and the kinematics of jets are included in terms of pseudorapidity $\eta$ of leading and next-to-leading jets. 
        
        To exploit the ISR characteristic of the ggH channel, we also included set of jet substructure variables $JS \equiv \{ n_{\text{track}}, \text{girth},$ $\text{broadening}, C_1^{\beta = 0.2}, \text{RMS}-p_T, \text{Pull-vector}\}$ for each jet. See Appendix \ref{app:jet substructure variables} for exact definitions of jet substructure variables. Therefore, the total sets of input variables including jet substructure, $J_{n_{\jet}}$, are defined as $J_0 \equiv S_0$, $J_1 \equiv S_1 \cup JS_1$ and $J_2 \equiv S_2 \cup JS_1 \cup JS_2$. 
        
        \subsection*{Training and categorization}
        The discriminators are trained separately for different $n_{\jet}$. One million signal and background events each are used for training (70\%) and validation (30\%). Three different machine discriminators are trained for comparison.
        
        After training, total data is categorized based on the output of the discriminators, or equivalently, by the signal efficiency. Boundaries are recursively chosen to maximize total expected significance ($S/\sqrt{B}$) as a figure of merit, while minimizing the number of categories until the gain from an additional boundary is less than 1\%.
            
        \subsection*{Absence of shape fitting}
        Typically, the next step involves fitting data to the $m_{\mu\mu}$ distribution to extract the number of events. However we will skip the extraction procedure and use the number of events by simulation directly because we do not have real observed data available for fitting. Fitting simulated data again would not be worth much since we already used large dataset. While our results rely on simulation, it is sufficient for comparing discrimination models with and without decorrelation, and assessing the impact of jet substructure variables featuring the properties of the jets associated with the dimuon system. Therefore after the categorization, we will compute the discovery significance by profile likelihood fitting to the $m_{\mu\mu}$ distribution.
            \section{\label{sec:discriminator} Discriminators}
        \subsection*{Boosted Decision Tree}
        The CMS collaboration used boosted decision tree (BDT) to categorize their ggH data. A tree-based model such as BDT is essentially a complex collection of cuts on training variables. This makes BDTs generally well-suited for tasks with theoretically well-motivated variables such as transverse momentum ($p_T$), pseudo rapidity ($\eta$) and so on. It is also easy to interpret. However, compared to network-based models like DNN, it lacks the ability of capturing complex non-linear relationships among training variables. Generally it is the reason why network based discriminators are preferred for more complex data analysis. Ironically, this intrinsic weakness of tree-based model would be profitable for making the resultant discriminator to be independent to $m_{\mu\mu}$.
        
        For comparison, we also performed BDT discrimination using the gradient-boosted tree provided by \textbf{SCIKIT LEARN} with Friedman mean squared error as the splitting criterion and deviance as the loss function. The details of trained model can be checked in Appendix \ref{app: BDT result}. Fig. \ref{fig:distortion_BDT} shows $m_{\mu\mu}$ distributions categorized by the output of BDT discriminators.
            \begin{figure*}[htb!]
                \centering
                \includegraphics[width=2\columnwidth]{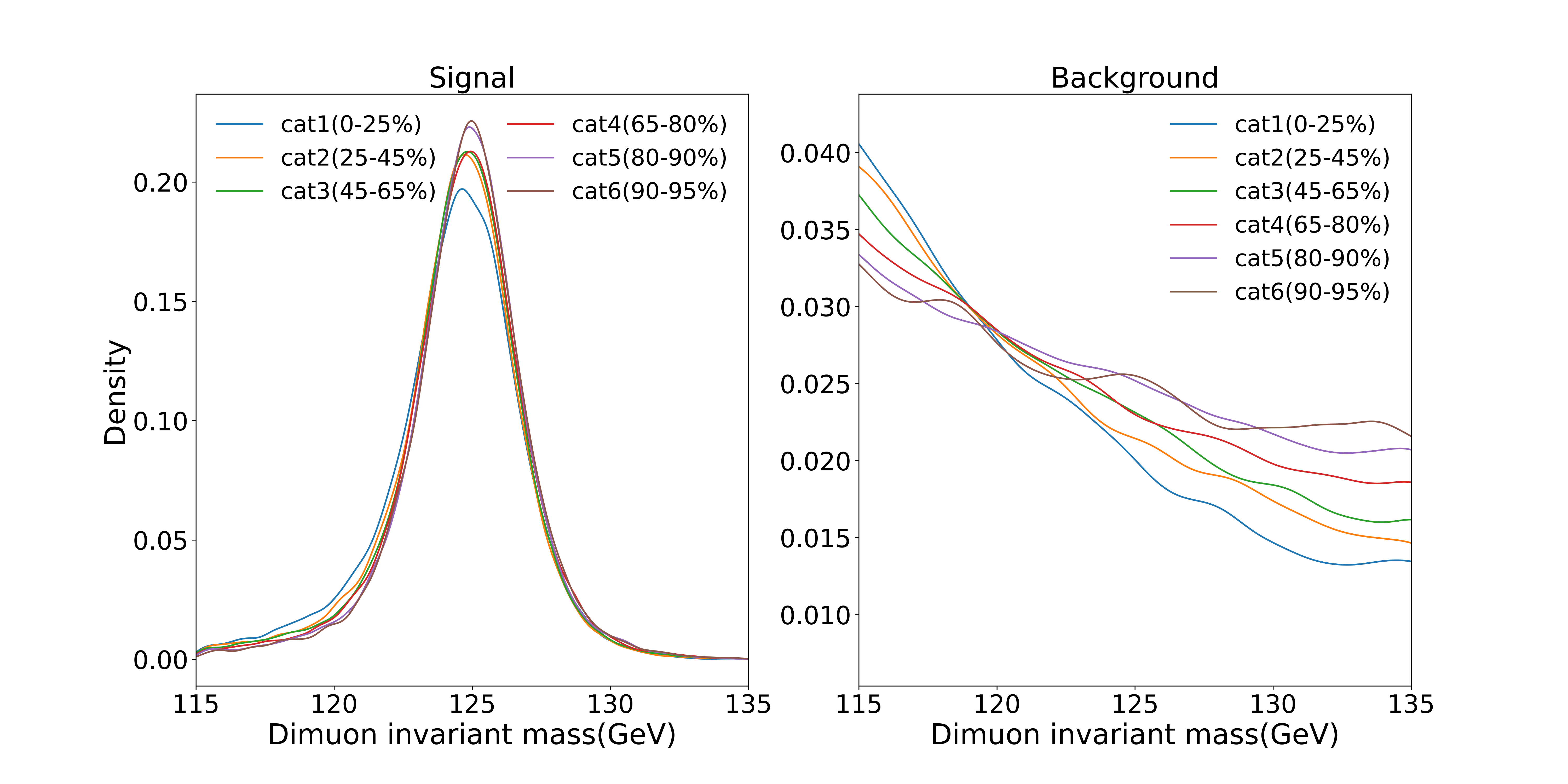}
                \caption{BDT; Normalized $m_{\mu\mu}$ distributions binned by optimal boundaries chosen by output of BDT discriminators trained without jet substructure variables ($S_{n_{\text{jet}}}$).}
                \label{fig:distortion_BDT}
            \end{figure*}
        The $m_{\mu\mu}$ distributions categorized by BDT are similar across the categories, implies that the choice of input variables having no or weak correlation with $m_{\mu\mu}$ worked relatively well for BDT. However, the non-zero correlation manifests itself in different slopes of background distributions. It is problematic for extracting the number of signal events. This issue was already recognized by the CMS collaboration. They used mass sideband events to constrain the background distribution with a modified fitting function having common shape and category specific slope, known as the `core-pdf' method. We tried to eliminate the correlation \textit{quantitatively} by including the distance correlation loss without a posteriori prescription. The definition of distance correlation is given in Appendix \ref{app:distance_correlation}. Unfortunately, the distance correlation is incompatible with BDT. The BDT requires decomposability of loss, but the distance correlation can not be defined locally for each point. On the other hand, the DNN can accommodate it, so we conducted DNN with and without decorrelation loss too. 
                    
        \subsection*{Deep Neural Network}
        The astonishing success of DNN is based on the ability to learn powerful hidden representations from training variables. However, due to the very feature, DNN can reconstruct the invariant mass even if it is not directly included in training variables. As a result, the distributions of categorized events can be largely distorted.
        
        We used \textbf{PyTorch} \cite{pytorch} to implement DNN and used \textbf{SCIKIT LEARN} \cite{Scikit-learn} for preprocessing the training data. The details of trained model can be checked in Appendix \ref{app: dnn result}. Fig. \ref{fig:distortion_DNN} shows the actual distortion by ordinary DNN discriminator. In categories with high signal purity, even the background distribution resembles the resonance peak, which can mimic signals and overestimate the signal counts. The severe distortion in DNN result implies that the DNN is capable of reconstructing $m_{\mu\mu}$ from input variables which have complex non-linear relationships with $m_{\mu\mu}$. It is not reliable to extract the number of events by fitting to $m_{\mu\mu}$ under such distortion.
            \begin{figure*}[htb!]
                \centering
                \includegraphics[width=2\columnwidth]{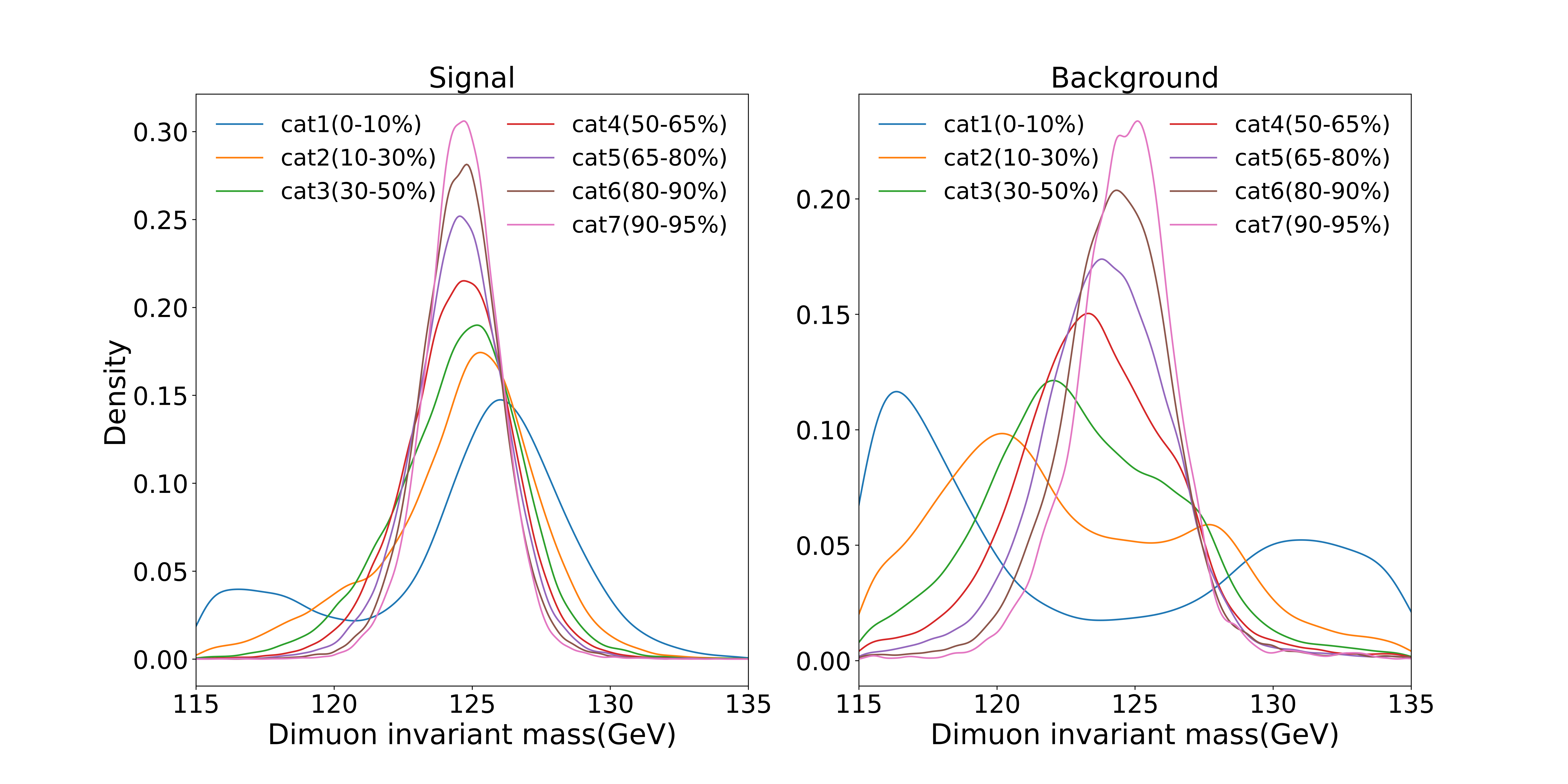}
                \caption{DNN; Normalized $m_{\mu\mu}$ distributions binned by optimal boundaries chosen by output of (ordinary) DNN discriminators trained without jet substructure variables ($S_{n_{\text{jet}}}$).}
                \label{fig:distortion_DNN}
            \end{figure*}
        In such cases, fitting a signal strength modifier directly to the output of discriminator (score) distribution might be a more tailored approach, but this approach often comes with significant drawbacks. This method is susceptible to larger systematic uncertainties that originate from two sources: the inherent uncertainties in Monte Carlo simulation and the uncertainties from training algorithms themselves. These uncertainties lead to a large variance in the resulting score distribution and make it difficult to extract reliable signal strengths.
            
        Consequently, fitting to the invariant mass distribution remains to be the preferred method, being simpler and more robust in most cases. As a single well-understood variable, the invariant mass is less prone to the uncertainties that distort the score distribution, and this makes the invariant mass fit a more reliable indicator for extracting signal strengths and further analysis. Therefore, a proper implementation of decorrelation is inevitable to achieve high precision.
                    
        \subsection*{Decorrelated DNN}
        For both BDT and DNN, the discrimination result is not statistically independent to $m_{\mu\mu}$. To achieve the independence, we exploit the flexibility of DNN. The network in DNN is an analytic function and it can handle wide variety of loss functions, which only requires to be differentiable. Therefore we introduce a `decorrelation loss' $\mathcal{L}_{\text{decor}}$ with a parameter $\lambda$ as $\mathcal{L}_{\text{tot}} \equiv \mathcal{L}_{\text{discrim}} + \lambda \mathcal{L}_{\text{decor}}$, where the discrimination loss $\mathcal{L}_{\text{discrim}}$ is the binary cross entropy. Here, $\mathcal{L}_{\text{discrim}}$ and $\mathcal{L}_{\text{decor}}$ are inherently conflicting. While $\mathcal{L}_{\text{discrim}}$ prioritizes distinguishing signal from background, $\mathcal{L}_{\text{decor}}$ aims to weaken the correlation between the discrimination score and $m_{\mu\mu}$. To achieve such decorrelation, we can increase the weight of $\mathcal{L}_{\text{decor}}$ by setting a sufficiently large $\lambda$ value, at the expense of some discrimination performance.

        The distance correlation ($\equiv \mathcal{R}$) is used as a measure of statistical dependence. We required the discrimination score to be independent to $m_{\mu\mu}$ separately for signal and background events, $\mathcal{L}_{\text{decor}} \equiv \mathcal{R}(m_{\mu\mu}, \text{score} | \text{signal}) + \mathcal{R}(m_{\mu\mu}, \text{score} | \text{background})$. We will call this DNN model with decorrelation loss as `decorrleated DNN' from now on.
        
        The decorrelated DNN is implemented with \textbf{PyTorch} again. We applied the decorrelation loss term every two epochs for calculation efficiency. Furthermore, since the distance matrix for the computation of distance correlation requires large memory, 1\% of every batch is randomly sampled to compute $\mathcal{R}$. To test the degree of decorrelation at the end of training, the distance correlation is calculated with sample of size 25,000 for signal and background respectively due to GPU memory limitations. Selected models are given in Appendix \ref{app: decor result}. As shown in Fig. \ref{fig:no_distortion}, the $m_{\mu\mu}$ distributions for the decorrelated DNN exhibit minimal distortion across categories. This indicates a successful reduction in the dependence between the discrimination score and the invariant mass. Consequently, the systematic errors arising from this dependence can be significantly mitigated.
            \begin{figure*}[hptb!]
                \centering
                \includegraphics[width=2\columnwidth]{hmumu/plots/categ_unlearning_cms.png}
                \caption{decorrelated DNN; Normalized $m_{\mu\mu}$ distributions binned by optimal boundaries chosen by output of decorrelated DNN discriminators trained without jet substructure variables ($S_{n_{\text{jet}}}$).}
                \label{fig:no_distortion}
            \end{figure*}

            \subsection{Comparison}
            As previously discussed, the decorrelation loss introduced in the decorrelated DNN inherently conflicts with the standard discrimination loss. This trade-off is crucial when comparing the three machine discriminators: BDT, ordinary DNN and decorrelated DNN. We can evaluate them based on two key properties: their discrimination performance (ability to distinguish signal from background) and their degree of decorrelation to the invariant mass $m_{\mu\mu}$.
            
\begin{figure}[hpbt!]
                     \includegraphics[width=\columnwidth]{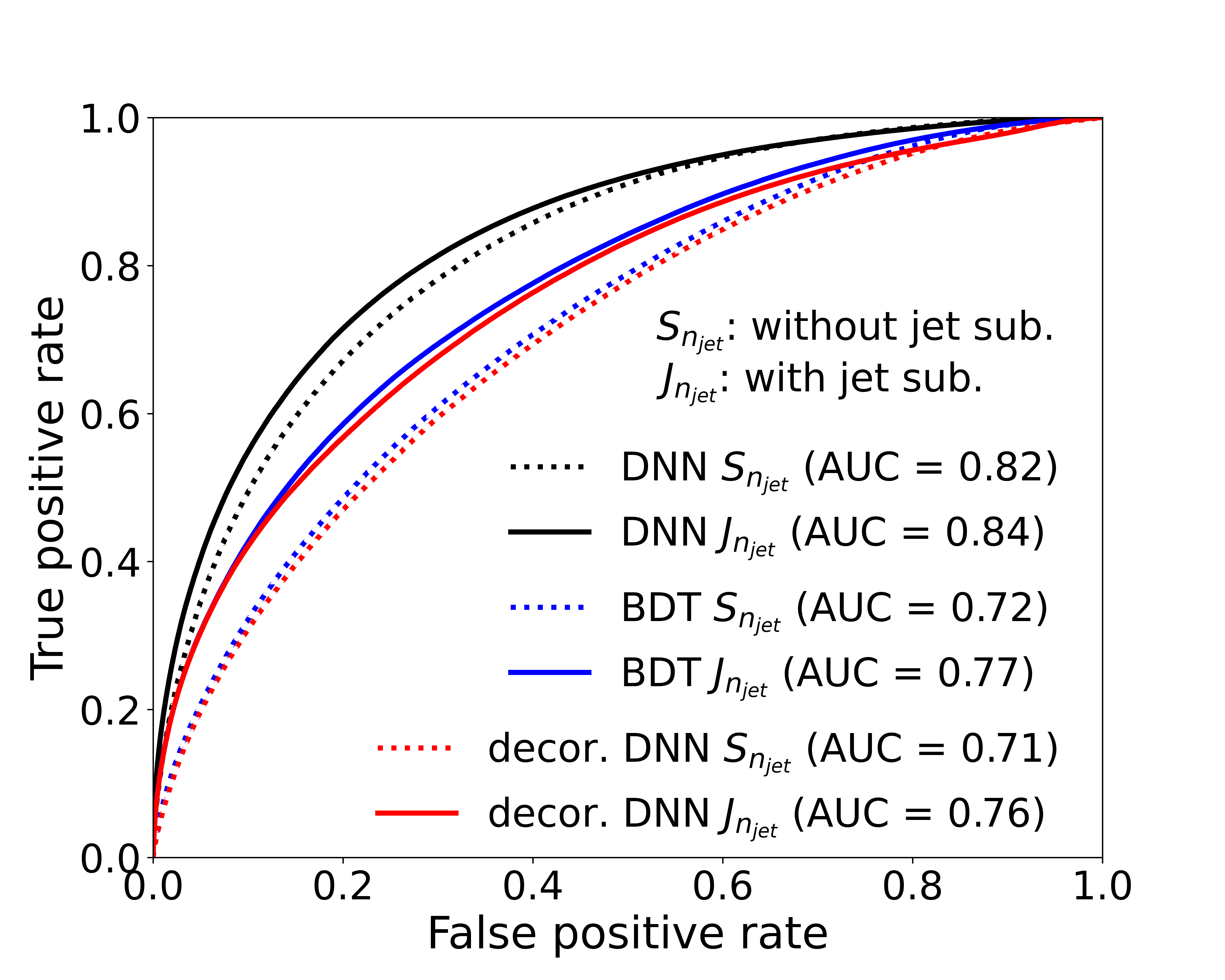}
                     \caption{Receiver operating characteristic (ROC) curve for three discrimination schemes - BDT, (ordinary) DNN and decorrelated DNN - with two different set of variables($S_{n_{\text{jet}}}$ and $J_{n_{\text{jet}}})$. The area under curve (AUC) is also given. Note that the true positive rate is equivalent to signal efficiency and the false positive rate is equivalent to (1-background rejection rate).}
                     \label{fig:ROC_curve}
                \end{figure}

            Fig. \ref{fig:ROC_curve} displays the Receiver Operating Characteristic (ROC) curve with Area Under the Curve (AUC) for each discriminator configuration. For all three discriminators, including jet substructure variables (solid lines) improves the discrimination performance of ggH channel. 
                
            The ordinary DNN (black line) achieves a substantially higher AUC, which is around 10\% higher, compared to BDT (blue line) and decorrelated DNN (red line). This is likely because the ordinary DNN can internally exploit $m_{\mu\mu}$ for discrimination, even though it was not explicitly included as a training variable. However, this high performance comes at a cost of the robustness of analysis related to $m_{\mu\mu}$ (Fig. \ref{fig:distortion_DNN}). 
            
            In contrast, the BDT and decorrelated DNN, which are inherently limited in their ability to reconstruct $m_{\mu\mu}$ internally, are trained to be insensitive to it. They result in the similar $m_{\mu\mu}$ distributions across score categories (Fig. \ref{fig:distortion_BDT} and \ref{fig:no_distortion}), but with lower AUCs compared to the ordinary DNN. While both exhibiting similar AUCs, the slight difference in performance between the BDT and decorrelated DNN indicates that there might still be some residual dependence on $m_{\mu\mu}$ in BDT.

            The degree of decorrelation between $m_{\mu\mu}$ and the output of discriminators can be seen directly from the distortion of invariant mass distribution categorized by the output of discriminators (Fig. \ref{fig:distortion_BDT}, \ref{fig:distortion_DNN}, \ref{fig:no_distortion}). It can also be quantitatively checked by a test of independence using asymptotic distribution of distance covariance $\mathcal{V}$ \cite{Sz_kely_2007}. The test statistic is 
                $$ V \equiv \frac{n \times \mathcal{V}^2(X,Y)}{S_2}$$
            where $n$ is sample size and 
                $$S_2\equiv \frac{1}{n^2}\sum_{k,l} |X_k - X_l| \frac{1}{n^2}\sum_{k,l} |Y_k - Y_l|$$ 
            The statistical independence would be rejected if 
                $$\sqrt{V} > \Phi^{-1}(1-\alpha/2)$$ 
            for selected significance level $\alpha$, where $\Phi$ is the Gaussian cumulative density function.
            
\begin{figure}[hbt!]
                    \centering
                    \includegraphics[width=\columnwidth]{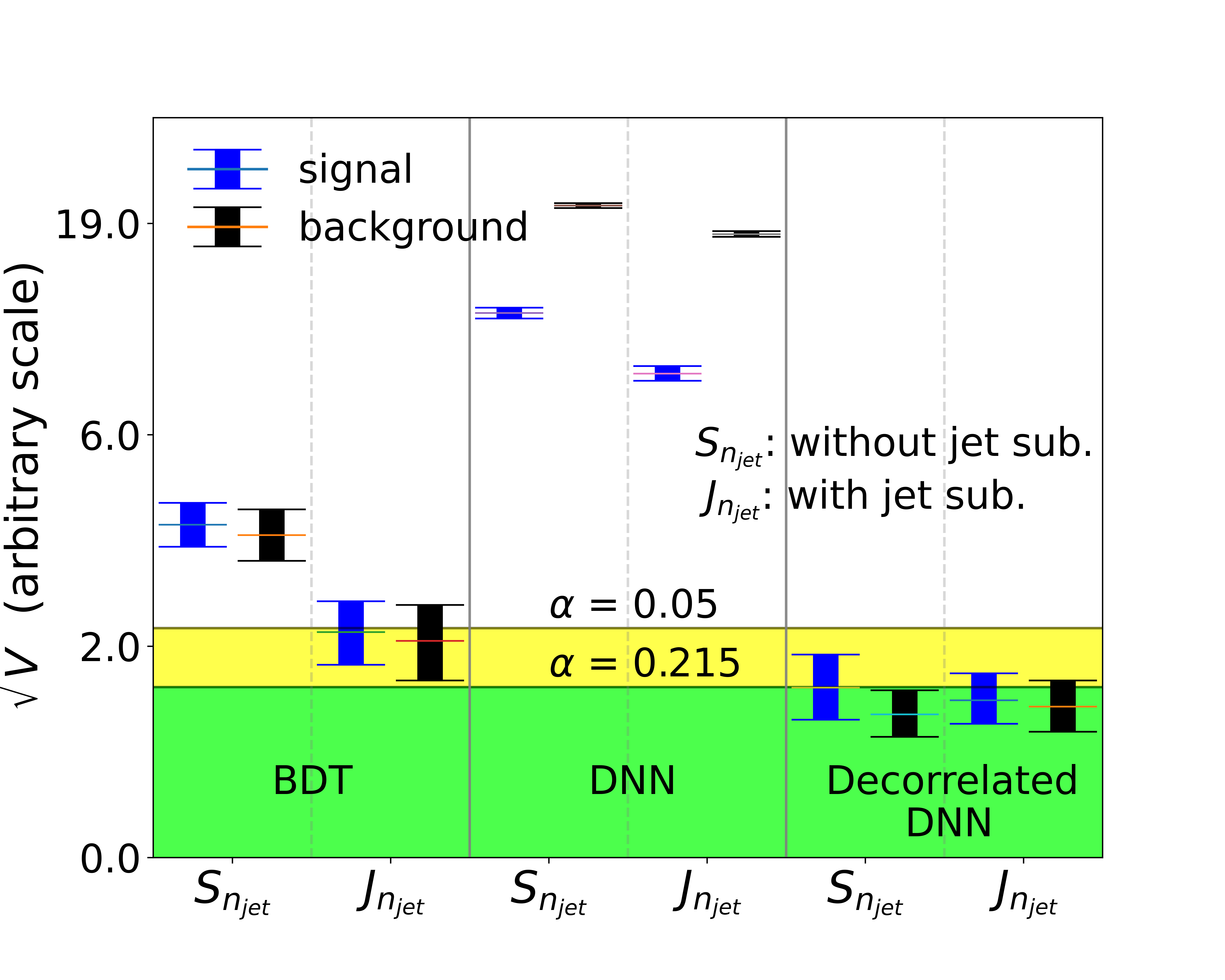}
                    \caption{The test statistic $\sqrt{V}$ of signal and background separately for various discriminators. The error bar comes from 1,000 iterations and sample size 10,000 was used. Two lines according to significance level $\alpha=0.05$  and $\alpha=0.215$ are given. Please note that the scale of y-axis is arbitrary.}
                    \label{fig:test result}
                \end{figure}

            Fig. \ref{fig:test result} shows the statistic $\sqrt{V}$ between $m_{\mu\mu}$ and the output of discriminator. Samples of size 10000 with 1000 iterations are used.
            It is obvious that $m_{\mu\mu}$ and the output of discriminators are enormously correlated for (ordinary) DNN. The independence hypothesis is always rejected for DNN case even under $\alpha < 0.05$. The situation is improved in the case of BDT but the uncontrolled impact of $m_{\mu\mu}$ still exist. The BDT trained with jet substructure variables barely pass the test up to $\alpha = 0.05$. On the other hand, the output of decorrelated DNN and $m_{\mu\mu}$ is independent up to $\alpha = 0.215$, which is the upper limit of significance level given in \cite{Sz_kely_2007}.
            
            It is interesting that the results become more independent for both BDT and DNN cases when the jet substructure variables are included for training. This is because jet substructure variables have discrimination power other than the $m_{\mu\mu}$ so that the distances of discrimination score have a more spread distribution. The decorrelated DNN result trained with $S_{n_{\text{jet}}}$ is already decorrelated to $m_{\mu\mu}$ and thus there is no further decorrelation by including jet substructure variables.
            
            In summary, it is verified that for all the machine discriminators, the discrimination performance is improved by including jet substructure variables. The gap between AUC of decorrelated DNN and those of other methods is due to the discriminatory power directly attributable to $m_{\mu\mu}$. At the same time, the decorrelation up to $\alpha = 0.215$ was achieved by our decorrelation method both choice of training variables, $S_{n_{\text{jet}}}$ and $J_{n_{\text{jet}}}$. It achieves sufficient independence from the invariant mass, for both the Higgs signal and the background simultaneously. The method has a significant potential for more robust Higgs boson to dimuon decay searches and precision measurements using the invariant mass projection.

            \subsection{Decorrelated DNN result in detail}
            Imposing distance correlation loss reduces the accuracy of decorrelated DNN but achieves the desired independence from $m_{\mu\mu}$. Consequently, the $m_{\mu\mu}$ can now be reliably used for analysis after the categorization process. We will now delve into a detailed examination of the categorization results of decorrelated DNN.

\begin{figure}[hbt!]
                    \subfloat[\textbf{without} jet substructure variables ($S_{n_{\text{jet}}}$) \label{subfig:categorization_decorrelated_cms}]{\includegraphics[width=\columnwidth]{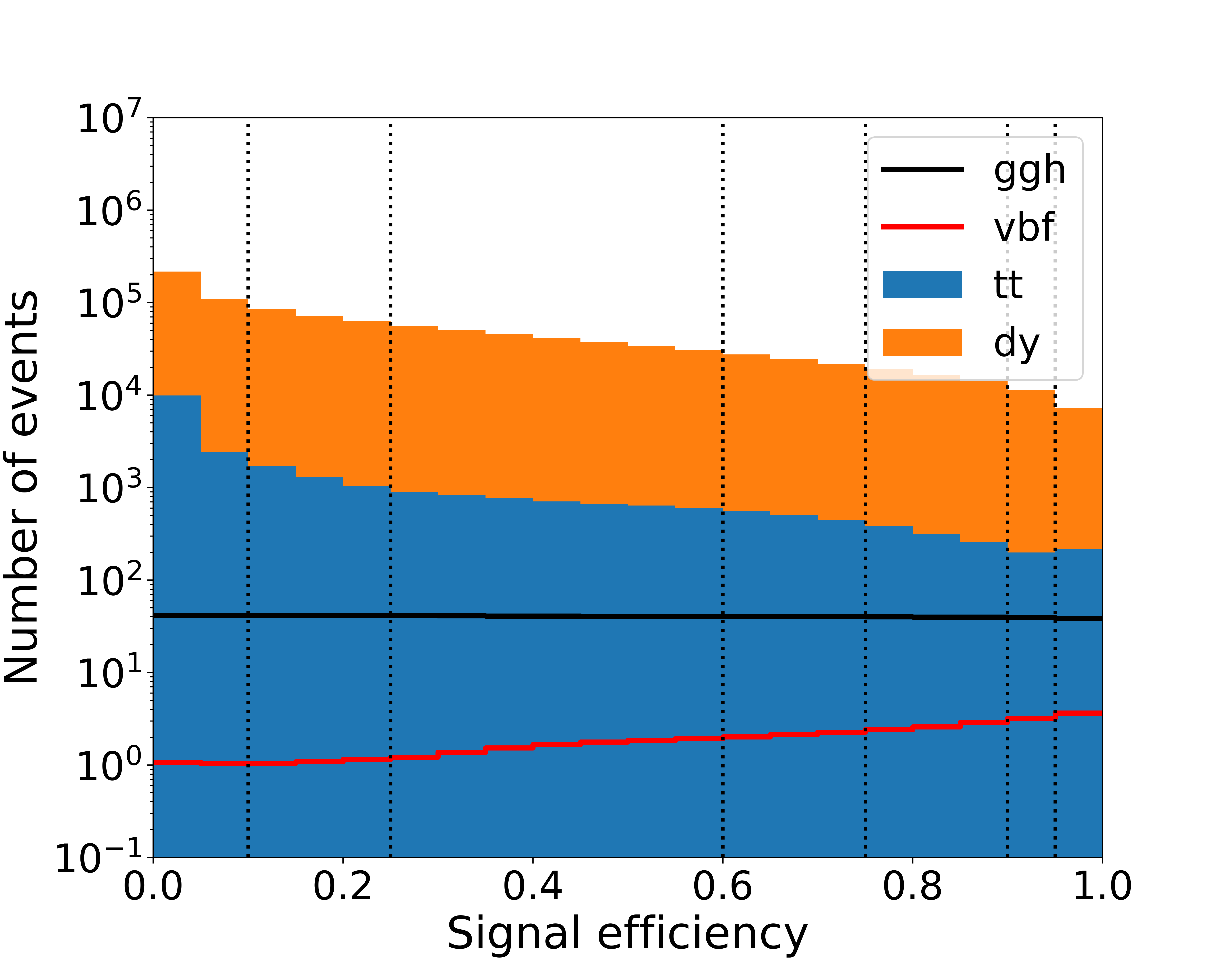}}
                         \hfill
                    \subfloat[\textbf{with} jet substructure variables ($J_{n_{\text{jet}}}$)\label{subfig:categorization_decorrelated_jetsub}]{\includegraphics[width=\columnwidth]{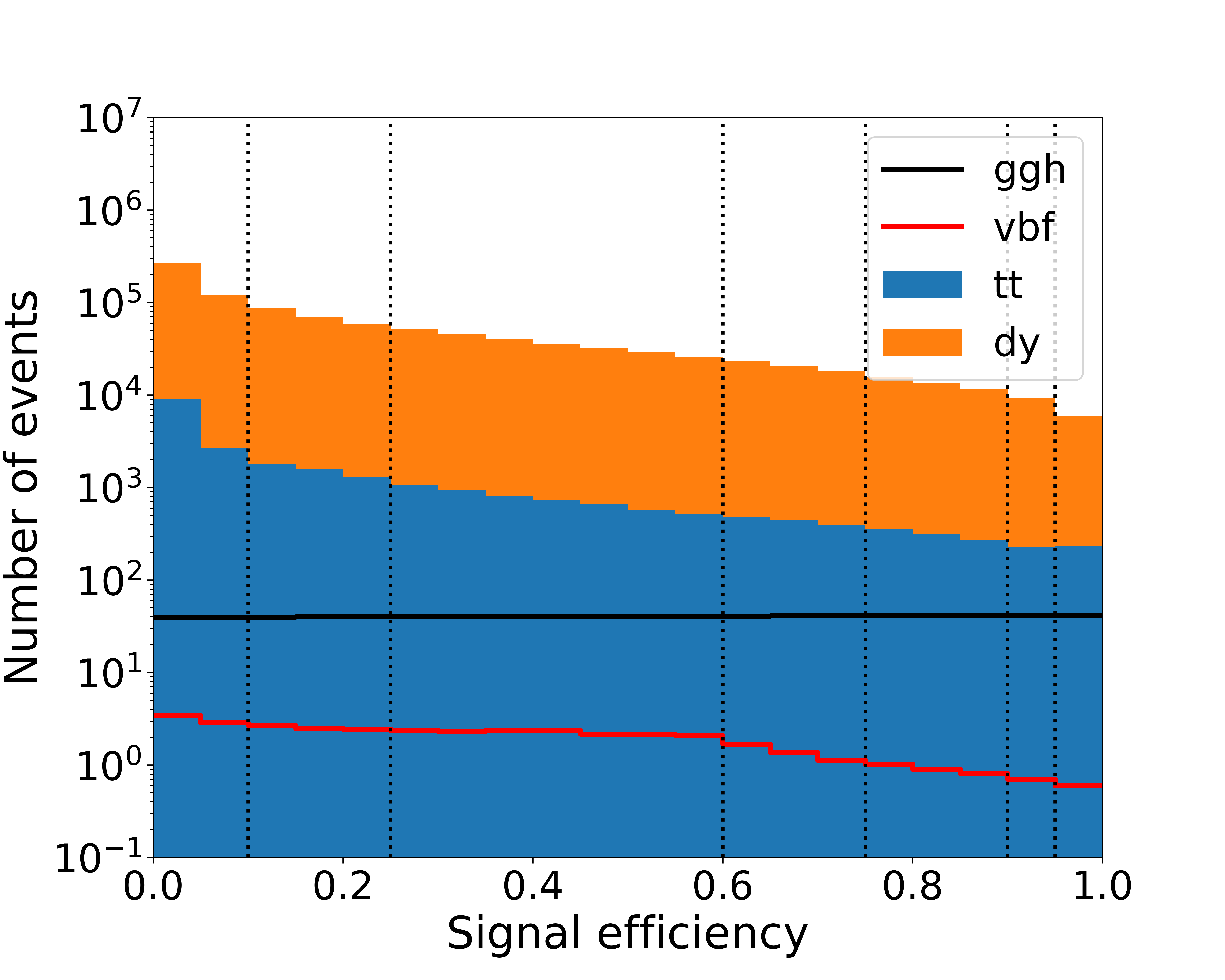}}
                    \caption{Number of events according to optimized bins according to signal efficiency. (a)/(b): For a model trained \textbf{without/with} jet substructure variables.}
                    \label{fig:categorization_decorrelated}
                \end{figure}

            Fig. \ref{fig:categorization_decorrelated} shows the number of events for each category and production channel. The number of signal and background is stacked respectively. The vertical dotted lines are the optimal boundaries. Exact number of signal events, background events, $S/\sqrt{B}$ and other information for each optimized category with and without jet substructure variables can be checked in App. \ref{app: detailed}.
            
             In Fig. \ref{subfig:categorization_decorrelated_cms}, the proportion of VBF events within the signal increases for categories with higher purity. On the other hand, the tendency of growing VBF portion is totally reversed for the discriminators trained with jet substructure variables. In Fig. \ref{subfig:categorization_decorrelated_jetsub}, the portion of VBF events are getting smaller for higher signal purity category. This reversal of VBF portion represents the improved discrimination power of ggH signals by exploiting jet substructure variables. This property allows for the reduction of errors arising from separate analysis of each production channel, ultimately improving the precision of the ggH channel measurement.

            \section{Discovery significance} \label{sec: discovery significance}
            We calculated the statistical significance using the profile likelihood ratio fit to the $m_{\mu\mu}$ distribution. We employed uniform binning of size $50$ MeV. To account for systematic errors, nuisance parameters with Gaussian priors were incorporated into the fit. Since the simulation was done for an arbitrary luminosity, the number of events are normalized to that of \cite{CMS:HiggsToDimuon}, $137\ $fb$^{-1}$. We used non-integers from the normalization directly to Poisson counts to avoid the subtle risk of getting the over- or under-estimated results from the rounding off. The problem of non-integers for Poisson distribution is not a genuine problem since it will be canceled off by calculating likelihood ratio \cite{cowan2011asymptotic}.
            
            We computed the statistical significance using both BDT and decorrelated DNN to compare their results. The ordinary DNN was excluded due to significant distortions in the cateogirzed $m_{\mu\mu}$ distribution. Please note that the significance from BDT is also computed under the assumption of statistical independence, which is not true actually. This distortion can give vastly wrong number of events if we fit to the $m_{\mu\mu}$, but our result does not reflect it since we used the simulation data directly rather than extraction from the fitting. The correct computation requires cumbersome combining procedure for significance of correlated categories. We can not predict the significance accurately since it largely depends on the prescriptions such as the `core-pdf' method used in \cite{CMS:HiggsToDimuon}.
            
            Table \ref{tab:merged_summary} shows the AUC and significance. Clearly, significance increases about 0.15 by exploiting ggH ISR characteristic. The slight difference of AUC and significance between the BDT and decorrelated DNN can be interpreted as uncontrolled impact of $m_{\mu\mu}$. It should be emphasized that the decorrelated DNN does not necessarily result in increased discovery significance. Decorrelation is an essential ingredient for the analysis with high precision rather than an optional choice.
            
              \begin{table}[hbt!]
                  \centering
                  \begin{tabular}{ c | c c | c c}
                    \multicolumn{5}{c}{~}\\
                    \hline
                    Machine & \multicolumn{2}{c|}{BDT} & \multicolumn{2}{c}{decorrelated DNN} \\
                    Variables & $S_{n_{\text{jet}}}$ & $J_{n_{\text{jet}}}$ & $S_{n_{\text{jet}}}$ & $J_{n_{\text{jet}}}$ \\
                \hline
                significance & (1.64) & (1.78) & 1.61 & 1.76 \\
                AUC & 0.72 & 0.77 & 0.71 & 0.76 \\
                \hline
                  \end{tabular}
                  \caption{Discovery significance Table CMS benchmark was 1.56$\sigma$ for ggH and 1.77 $\sigma$ for VBF. The numbers in the parenthesis are obtained without sufficient decorrelation and should be interpreted cautiously.}
                  \label{tab:merged_summary}
              \end{table}

            \section{Summary}
            We applied decorrelation using distance correlation to the Higgs to dimuon discovery from ggH channel. The decorrelation between output of discriminator and $m_{\mu\mu}$ allows us to categorize the events without any systematic error from $m_{\mu\mu}$ correlations. The decorrelated DNN achieved the independence for both signal and background events simultaneously up to significance level $\alpha = 0.215$. The jet substructure variables are helpful for improving discovery significance of ggH channel and the significance was improved from $1.61$ to $1.76$.
            
            Decorrelation can be helpful in general for both precision and discovery as it can preserve the variable of interest-distribution of events after the cut and get rid of systematic errors from distortion in the process of categorization. In addition, all the variables can be used for categorization without any concern of correlation, simplifying the feature selection process. Decorrelation is an essential ingredient for the analysis with high precision. Application of decorreation to the VBF channel and decorrelation using tree-based discriminators can be done in following research.

            \section{Acknowledgements}
            We thank Wonsang Cho for collaboration at an initial stage of the work and for discussion and comments.
            This work is supported by NRF of Korea grant, No. 0426-20230001 and also by the use of computing resources funded by the NRF, Grant No. 2017R1C1B2011048.

            \appendix
            \section{Training variables} \label{app: variables}
            The set of training variables for events with each number of jets $n_{\text{jet}}$ will be summarized. They were carefully selected to exclude the $m_{\mu\mu}$ information.
    
        \subsection{Without jet substructure variables} \label{app: cms variables}
        The set of training variables $S_{n_{\text{jet}}}$ without jet substructure variables, which is the same as the one used by CMS collaboration, are
        \begin{align*}
            S_0 \equiv 
            &~\{ p_T^{\mu\mu}, y_{\mu\mu}, \phi_{CS}, \cos \theta_{CS}, \eta(\mu_1),  \eta(\mu_2)\\
            &~, \frac{p_T^{\mu_1}}{m_{\mu\mu}} , \frac{p_T^{\mu_2}}{m_{\mu\mu}} \} \\
            S_1 \equiv
            &~ S_0 \cup \{p_T(j_1), \eta(j_1), \Delta R(\mu\mu, j_1) \} \\
            S_2 \equiv 
            &~S_1 \cup\{ p_T(j_2), \eta(j_2), m_{jj}, \Delta\eta_{j_1,j_2}, \Delta\phi_{j_1,j_2} \\
            &~, \Delta\eta_{\mu\mu,j_1}, \Delta\eta_{\mu\mu,j_2}, \Delta\phi_{\mu\mu,j_1}, \Delta\phi_{\mu\mu,j_2}\\
            &~ , z_*, n \}
        \end{align*} 
        where $p_T$ is transverse momentum, $y$ is rapidity, $\theta, \phi$ are zenith and azimuthal angle respectively, $R$ is the absolute distance in $\eta, \phi$ space, $z* \equiv \frac{y_{\mu\mu}-(y_{j_1} + y_{j_2}) / 2}{|y_{j_1} - y_{j_2}|}$ is the Zeppendfeld variable and $n$ is the number of jets with $p_T > 25$GeV, $|\eta| < 4.7$.
        The index CS means those computed in the dimuon Collins-Soper rest frame, $j_{1,2}$ respectively denotes the highest and second highest $p_T$ jet, $\mu_1, \mu_2$ denote two muons, $\mu\mu$ and $jj$ denote dimuon and dijet of $j_1, j_2$ system respectively.

        \subsection{Jet substructure variables}\label{app:jet substructure variables}
        The jet substructure variables are generally defined as weighted moment sum of constituent transverse momenta normalized by transverse momentum of the jet to capture the fatness of gluon jet. For each jet, we used the number of charged tracks in the jet $n_{\text{track}}$ \cite{Gallicchio_2011}, girth (linear radial moment)  \cite{Gallicchio_2011, girth}, broadening \cite{broadening}, Energy-energy correlation(EEC) $C^{\beta = 0.2}_1$ with $\beta = 0.2$   \cite{Larkoski_2013,Dasgupta_2013}, RMS-$p_T$ \cite{Gallicchio_2011}, Pull-vector \cite{pull}. 
            \begin{align*}
                \textrm{girth } &
                G = \frac{1}{p^{\text{jet}}_T} \displaystyle\sum_{i \in \textrm{jet}} p^i_T |\Delta \Vec{r}_i| \\
                \textrm{broadening } 
                &B = \frac{1}{\sum_i |\Vec{p \ }^i|} \displaystyle\sum | \Vec{p \ }^i \times \Hat{p}^{\textrm{jet}}|\\
                &~~ = \frac{1}{\sum_i |\Vec{p \ }^i|} \displaystyle\sum |\Vec{k \ }^i_T| \\
                \textrm{EEC } &C^{\beta}_1 
                = \frac{1}{(\sum_i p^i_T)^2} \displaystyle\sum_{i < j} p^i_T p^j_T (\Delta R_{ij})^{\beta} \\
                \textrm{RMS-$p_T$ } &\sqrt{<p^2_T>} 
                = \frac{1}{p^{\textrm{jet}}_T} \sqrt{\frac{1}{n_{\textrm{tk}}} \displaystyle\sum_i (p^i_T)^2} \\
                \textrm{Pull-vector } &\Vec{v}_p 
                = \frac{1}{p^{\textrm{jet}}_T} \displaystyle\sum_i p^i_T |\Delta \Vec{r \ }^i| \Delta\Vec{r \ }^i
            \end{align*}

            \section{Distance correlation}\label{app:distance_correlation}
            The distance covariance ($\mcV$) and distance correlation ($\mcR$) are measures of both linear and non-linear correlation between two random variables/vectors \cite{Sz_kely_2007}. The distance covariance is defined as an weighted $L_2$ norm between the joint characteristic function and the product of marginal characteristic functions ($\phi$).
                $$\mathcal{V}^2(X,Y) 
                    \equiv \int dw(t,s) |\phi_{XY}(t,s) - \phi_X(t)\phi_Y(s)|^2$$
            where the weight $dw$ is chosen for $\mcV$ to be scale equivariant and therefore for the distance correlation (which is to be defined soon) to be scale invariant. Moreover, to capture the non-linear correlation, $dw$ is chosen to be nonintegrable. It is clear that $X$ and $Y$ are independent if and only if $\mathcal{V}^2(X,Y) = 0$ by the property of characteristic function. The distance correlation is a distance variance normalized into a value in $[0, 1]$.
                $$ \mathcal{R}^2(X,Y) \equiv \frac{ \mathcal{V}^2(X,Y) }{\sqrt{\mathcal{V}^2(X,X) \mathcal{V}^2(Y,Y)}}$$

            The distance covariance and distance correlation are also well defined in the sample space. For notational convenience, define doubly centered matrix as $\Bar{A}_{kl} \equiv A_{kl} - A_{k\cdot} -A_{\cdot l} + A_{\cdot \cdot}$ where $A_{kl} = |X_k - X_l|$ and $\cdot$ represent the mean along that index, e.g. for sample size $n$, $A_{k\cdot} \equiv \frac{1}{n}\sum_l A_{kl}$. Similarly define $\Bar{B}$ for $Y$. Then the empirical distance covariance is defined as 
                \begin{align*}
                    \mathcal{V}^2_n(X,Y) 
                    &\equiv \sum_{k,l} \Bar{A}_{kl} \Bar{B}_{kl} = S_1 + S_2 - 2S_3
                \end{align*}
            where 
                \begin{align*}
                    S_1 & \equiv \frac{1}{n^2} \sum_{k,l} A_{kl}B_{kl} \\
                    S_2 & \equiv \frac{1}{n^2} \sum_{k,l} A_{kl} \frac{1}{n^2} \sum_{k,l}B_{k,l} = A_{\cdot \cdot} B_{\cdot \cdot} \\
                    S_3 & \equiv \frac{1}{n^3} \sum_{k,l,m} A_{kl}B_{km}
                \end{align*}
            The empirical distance correlation is defined by the above definition of empirical distance covariance   
                $$\mathcal{R}^2_n(X,Y) \equiv \frac{\mathcal{V}^2_n(X,Y)}{\sqrt{\mathcal{V}^2_n(X,X)\mathcal{V}^2_n(Y,Y)}}$$
            The consistency of definitions in population and sample space can be checked in \cite{Sz_kely_2007} and the differentiability comes directly from the definition above. Please note that we used empirical distance covariance and correlation without subscript $n$ since there is no ambiguity.
                    
            \section{Discrimination models in detail}\label{app:classification results}
        \subsection{BDT} \label{app: BDT result}
        A minimum sample size for a node to be split (min\_samples\_split) is 1\% to regularize the over-fitting. We used stochastic gradient boosting with $70\%$ of sample for each tree. Table \ref{tab:machine_BDT} shows the resultant (hyper-)parameters and accuracy for each machine. The minimum sample size after the splitting is once more limited by min\_samples\_leaf (MSL).
            \begin{table}[htb!]
                \centering
                \begin{tabular}{|c|c c c | c c |}
                    \hline
                    & \multicolumn{3}{c|}{w/o jet sub.} & \multicolumn{2}{c|}{w/ jet sub.}\\
                    Variables & $S_0$ & $S_1$ & $S_2$ & $J_1$ & $J_2$ \\ \hline
                    \hline
                    $n_{\text{tree}}$  & 80    & 1500  & 1000  &  1500      & 1000      \\
                    max depth           & 3     & 5     & 7     &  7         & 9       \\
                    learning rate           & 0.4   & 0.2   & 0.2   &  0.1       & 0.1      \\
                    MSL        & 0.001 & 0.01  & 0.02  &  0.001     & 0.001 \\
                    \hline
                    accuracy   & 0.66  & 0.65  & 0.66  &  0.72  &   0.73 \\
                    \hline
                \end{tabular}
                \caption{(hyper-)Parameters and accuracy of BDT. MSL stands for min\_samples\_leaf.}
                \label{tab:machine_BDT}
            \end{table}

        \subsection{(Ordinary) DNN}\label{app: dnn result}
        Table \ref{tab:DNN_new} shows the resultant (hyper-)parameters and accuracy for each ordinary DNN machine. 256 nodes per layer and 100 iterations of batch are commonly used. 
            
        \subsection{Decorrelated DNN}\label{app: decor result}
        Table \ref{tab:Disco DNN} shows the resultant (hyper-)parameters and accuracy for each decorrelated DNN machine. 128 nodes per layers and 100 batches are commonly used. In this analysis, we adopted a relative multiplier value of $\lambda = 8 \times 10^{-3}$. During the training, we dynamically adjust the relative multiplier $\lambda$. If the $\mathcal{L}_{\text{decor}}$ exceeded a threshold of $10^{-3}$, $\lambda$ was magnified by a factor of 100. Since the $\mathcal{L}_{\text{decor}}$ barely touch the threshold, $\lambda$ remained around $8 \times 10^{-1}$ for most of the training process. The the sum of distance correlations of signal and background $\mathcal{L}_{\text{decor}}$ is larger than the value of distance correlation computed for the test sample (**) since it is computed with smaller sample.
        \begin{table}[htb!]
                \centering
                \begin{tabular}{| c| ccc | cc |}
                    \hline
                    & \multicolumn{3}{c|}{w/o jet sub.}& \multicolumn{2}{c|}{w/ jet sub.}  \\
                    Variables  & $S_0$ & $S_1$ & $S_2$ & $J_1$ & $J_2$ \\ \hline
                    $n_{\text{layers}}$   & \multicolumn{3}{c|}{2} &  \multicolumn{2}{c|}{3}     \\
                    learning rate   & 0.007 & 0.01 & 0.003 & 0.04 & 0.005      \\
                    epochs          & 100 & 300 & 400 & 300 & 300     \\ \hline
                    accuracy        & 0.76 & 0.75 & 0.75 & 0.79 & 0.79 \\ \hline
                    $\mathcal{R}$(sig) & 0.35 & 0.39 & 0.33 & 0.32 & 0.24\\
                    $\mathcal{R}$(bkg) & 0.31 & 0.31 & 0.30 & 0.28 & 0.18 \\
                    \hline
                \end{tabular}
                \caption{(hyper-)Parameters and accuracy of DNN. The resultant distance correlation $\mathcal{R}$ the distance correlation $\mathcal{R}$ is calculated with 25,000 events each from the test sample.}
                \label{tab:DNN_new}
            \end{table}
            
            \begin{table}[htb!]
                \centering
                \begin{tabular}{|c| c c c | c c |}
                    \hline
                    & \multicolumn{3}{c|}{w/o jet sub.}& \multicolumn{2}{c|}{w/ jet sub.}  \\
                    Variables  & $S_0$ & $S_1$ & $S_2$ & $J_1$ & $J_2$ \\ \hline
                    $n_{\text{layers}}$   & 3 & 3 & 6 &  3 & 6      \\
                    learning rate  & 0.0004 & 0.001 & 0.01 &  0.001 & 0.001      \\
                    epochs          & 30 & 40 & 60 & 60 & 60     \\ \hline
                    $\mathcal{L}_{\text{decor}}$*  & 0.046 & 0.047 & 0.044 & 0.037 & 0.038 \\
                    accuracy         & 0.65 & 0.64 & 0.65 & 0.71 & 0.72 \\ \hline
                    $\mathcal{R}$(sig)** & 0.006 & 0.009 & 0.011 & 0.008 & 0.008\\
                    $\mathcal{R}$(bkg)** & 0.019 & 0.018 & 0.022 & 0.010 & 0.010 \\
                    \hline
                \end{tabular}
                \caption{Resultant model (hyper-)parameters and accuracy. *: $\mathcal{L}_{\text{decor}}$ is the sum of distance correlation loss between score and $m_{\mu\mu}$ respectively for signal and background. 1\% of every batch is ampled to calculate $\mathcal{L}_{\text{decor}}$. **: The distance correlation $\mathcal{R}$s are calculated with 25,000 events each from the test sample.}
                \label{tab:Disco DNN}
            \end{table}

                \section{Tables of decorrelated DNN result in detail}
                \label{app: detailed}
                Table \ref{tab:categorization_decorrelated_cms}, \ref{tab:categorization_decorrelated_jetsub} shows the number of signal, portion of each production channel, number of background, $S/(S+B)$ and $S/\sqrt{B}$ for optimized categories without and with jet substructure variables.
                
                \begin{table}[hbt!]           
                    \centering
                    \begin{tabular}{c | c | c | c | c | c | c}
                        \hline
                        Signal & Sig & ggH & VBF & Bkg & $\frac{S}{S+B}$& $\frac{S}{\sqrt{B}}$ \\
                        efficiency& & (\%) & (\%) & & (\%) &   \\
                        \hline
                        0-25\% & 212.45 & 97.5 & 2.5 & 545713 & 0.04 & 0.29 \\
                        25-45\% & 169.93 & 96.6 & 3.4 & 193690 & 0.09 & 0.39 \\
                        45-75\%& 170.17 & 95.6 & 4.4 & 130068 & 0.13 & 0.47\\
                        75-90\% & 212.65 & 94.2 & 5.8 & 96033 & 0.22 & 0.69\\
                        90-95\% & 42.65 & 92.5 & 7.5 & 11317 & 0.38 & 0.40\\
                        95-100\% & 42.41 & 91.3 & 8.6 & 7254 & 0.58 & 0.50\\
                        \hline
                    \end{tabular}
                    \caption{Categorization for decorrelated DNN \textbf{without} jet substructure variables.}
                    \label{tab:categorization_decorrelated_cms}
                \end{table}
    
                \begin{table}[hbt!]
                    \centering
                \begin{tabular}{c | c | c | c | c | c | c}
                    \hline
                    Signal & Sig & ggH & VBF & Bkg & $\frac{S}{S+B}$& $\frac{S}{\sqrt{B}}$ \\
                        efficiency& & (\%) & (\%) & & (\%) &   \\
                    \hline
                    0-10\% & 85.09 & 92.6 & 7.4 & 389125 & 0.02 & 0.14 \\
                    10-25\% & 127.42  & 94.0 & 6.0 & 216714 & 0.06 & 0.27 \\
                    25-60\%& 297.62  & 94.7 & 5.3 & 260441 & 0.11 & 0.58\\
                    60-75\% & 127.66  & 96.7 & 3.3 & 61487 & 0.21 & 0.51\\
                    75-90\% & 127.53 & 97.9 & 2.1 & 40971 & 0.31 & 0.63\\
                    90-95\% & 42.52 & 98.3 & 1.7 & 9348 & 0.45 & 0.44\\
                    95-100\% & 42.41 & 98.6 & 1.4 & 5933 & 0.71 & 0.55\\
                    \hline
                \end{tabular}
                \caption{Categorization for decorrelated DNN \textbf{with} jet substructure variables.}
                \label{tab:categorization_decorrelated_jetsub}
                \end{table}
                
\bibliography{main}

\end{document}